\documentclass[a4paper]{article}

\usepackage{amsmath,graphicx,amssymb,fancyhdr,amsthm,enumerate,textcomp} 
\usepackage[usenames]{color}

\theoremstyle{definition}

\theoremstyle{remark}  
  
\def\beq{\begin{eqnarray}}  
\def\eeq{\end{eqnarray}}  
\def\bsp{\begin{split}}  
\def\esp{\end{split}}

\def\Ric{\mathrm{Ric}}

\begin{document}   
   
\title{\Large\textbf{Opening Pandora's box: \\ Kundt solutions to Conformal Killing Gravity}}  
\author{{\large{Sigbj\o rn Hervik$^\clubsuit$ and Edmar G. Pantohan$^\diamondsuit$}    }
 \vspace{0.3cm} \\
{\small $^\clubsuit$ Department of Mathematics and Physics,
 University of Stavanger, Norway}\\
 {\small $^\diamondsuit$Department of Physics, Caraga State University, Philippines}\\
{\small \texttt{sigbjorn.hervik@uis.no, ~e.g.pantohan@gmail.com} }}
\date{\today}
\maketitle
\pagestyle{fancy}
\fancyhead{} 
\fancyhead[C]{Hervik and Pantohan}
\fancyhead[L,OR]{\thepage} \fancyhead[OC]{Kundt solutions to Conformal Killing Gravity}
\fancyfoot{} 

\begin{abstract}
We consider Kundt solutions to vacuum Conformal Killing Gravity (CKG) proposed by Harada \cite{Harada23} and find numerous solutions in four dimensions and in higher dimensions. In CKG theory the cosmological constant appears as an integration constant, and hence, is naturally embedded in the theory. However, by considering Kundt solutions to CKG we seemingly open Pandora's box: We find so many solutions so a selection criterium is needed in CKG to explain the current accelerated expansion of the universe. 
\end{abstract} 

\section{Introduction}
General relativity has been around for more than a hundred years and has had tremendous success. It has exposed some of the deepest secrets of the Universe, among them black holes, the Big Bang theory and the cosmic expansion of the universe. However, in spite of its success, there are a some properties of the Universe which are still elusive. For example, what is Dark matter, and what is the origin of Dark energy?  

Many alternative theories of gravity has been proposed to explain the dark sector of the universe, for example, models with exotic matter, higher-dimensional models, or higher-curvature theories of gravity. Recently, Harada \cite{Harada23} proposed a third order theory of gravity, where the cosmological constant appears as an integration constant and thus naturally appears in the solutions of the theory. In Harada's theory of gravity the equations are given in terms of the following totally symmetric tensors: 
\beq
H_{\alpha\mu\nu}&=&\frac{8\pi G}{c^4} T_{\alpha\mu\nu}, \\
H_{\alpha\mu\nu}&:=&\nabla_{(\alpha}R_{\mu\nu)}-\frac 13g_{(\mu\nu}\nabla_{\alpha)}R,\\
T_{\alpha\mu\nu}&:=&\nabla_{(\alpha}T_{\mu\nu)}-\frac 16g_{(\mu\nu}\nabla_{\alpha)}T,
\eeq
where $R_{\mu\nu}$ is the Ricci tensor and $T_{\mu\nu}$ is the energy-momentum tensor. 
Soon afterwards, Mantica and Molinari \cite{Mantica:2023} formulated Harada's theory interpreting it as a Conformal Killing Gravity (CKG) in the following sense: Harada's theory can be equivalently formulated as modifying the field equations using the existence of a conformal Killing tensor $K_{\mu\nu}$ so that
\beq
R_{\mu\nu}-\frac 12Rg_{\mu\nu}=\frac{8\pi G}{c^4}T_{\mu\nu}+K_{\mu\nu}.
\eeq
Since Harada's original proposal there have been several papers considering solutions to CKG, see e.g. \cite{Harada23-2,Barnes24,Mantica24,Barnes,Clement,Junior,Gurses}. 

Here, we will consider vacuum solutions of the CKG, i.e., $H_{\alpha\mu\nu}=0$. Furthermore, assuming the Ricci scalar is constant, $\nabla_{\alpha}R=0$, and the Harada vacuum equations imply:
\[ H_{\alpha\mu\nu}=\nabla_{(\alpha}R_{\mu\nu)}=0.\]
This immediately implies that solutions to the vacuum Conformal Killing Gravity include the following classes:
\begin{enumerate}
    \item {\bf Einstein metrics:} $R_{\mu\nu}=\lambda g_{\mu\nu}$. Examples in this class are the inflationary de Sitter spaces, and also the Anti-de Sitter spaces. All space-times being the solution of the vacuum Einstein equations, or Einstein equations with a cosmological constant, positive or negative, are in this class. 
    \item {\bf Symmetric spaces:} $\nabla_\mu {\rm Riem} =0$. Examples here are the direct product spaces (A)dS${}_2\times M_2$, where $M_2$ is any 2-dimensional maximally symmetric space, and  (A)dS${}_3\times \mathbb{R}$. 
    \item {\bf Ricci-symmetric spaces:} $\nabla_\alpha R_{\mu\nu}=0$. Examples in this category (which are not symmetric, nor Einstein) can be found in the Kundt class of metrics. For example, the general class of $pp$-waves solutions to CKG-vacuum found by Barnes \cite{Barnes} is in this class. 
\end{enumerate}
Hence, it is clear that such metrics include a wide variety of solutions. In what follows, we will provide with even more  solutions to the CKG vacuum equations which are not in the above-mentioned classes. 

In this paper we will look for solutions of the Kundt family of metrics, which in general can we written \cite{kundt}
\beq
2du\left(dv+H(u,v,x^k)du+W_i(u,v,x^k)dx^i\right)+h_{ij}(u,x^k)dx^idx^j,
\eeq
in particular those that have all constant scalar invariants \cite{CSI} (and hence, $\nabla_{\alpha}R=0$). This thus implies the CKG-vacuum equation is equivalent to $\nabla_{(\alpha}R_{\mu\nu)}=0$ and that the Ricci tensor is in fact a Killing tensor. Although the Einstein metrics, Symmetric and Ricci-symmetric cases of solutions is quite rich in itself, we will point out that there are numerous additional solutions, not belonging to the above classes. 

In particular, we will consider the 4-dimensional case but we will also consider higher-dimensional solutions to $\nabla_{(\alpha}R_{\mu\nu)}=0$. In the Kundt family, there are several solutions that are worth pointing out. One is the generalised Kaigorodov solution \cite{Kaigorodov,solvm2} which is an Einstein space, in addition to being space-time homogeneous (thus belongs to the 'trivial' Einstein class of solutions of the vacuum CKG-equations). The other, more interesting case, is the generalised Defrise solutions \cite{Defrise,solvm2}. These are not Einstein, but are space-time homogeneous. Especially the Defrise solutions play a role in finding other solutions. 

Kundt solutions to CKG have been considered before, for example, Barnes \cite{Barnes} investigated pp-wave solutions. More recently, G\"{u}rses, Heydarzade and \c{S}ent\"{u}rk \cite{Gurses} considered some Kundt metrics in CKG gravity. As pointed out, the Kundt assumption implies that the CKG-equations become a linear 3rd order partial differential equation in the function $H$. All the cases considered here can be interpreted as a wave-type of metric in a certain background. The background metric, $g_B$, are all either Einstein, or symmetric. Therefore, the Ricci metric of the background, $\Ric_B$ has particular simple form. By generalising to include a function $H$, the Ricci tensor gets the form: 
\[ \Ric = R_{11}{(\omega}^1)^2+\Ric_B\]
where $\omega^1$ is a null one-form. In terms of the algebraic classification of tensors \cite{MCPP}, the Ricci tensor $\Ric_B$ is of algebraic type D, while $\Ric$ is of type II. The solution to the linear equation of $H$ can typically be written as a sum of several terms: 
\[ H=H_{\text{Defrise}}+H_{{K}}+H_{\text{Kaigorodov}}+H_B\]
Both of the latter terms, the Kaigorodov term and the term $H_B$, preserve the Ricci metric and do not contribute to the component $R_{11}$. These solve the 2-order homogeneous equation stemming from the $R_{11}=0$ requirement. Due to its geometric significance, we have explicitly written the Kaigorodov term in the solutions below. The Defrise solution 
$H_{\text{Defrise}}$ is also of geometric significance and contributes both to the Ricci component $R_{11}$ and to $\nabla_\alpha R_{\mu\nu}$, but solves the 3rd-order differential equation $\nabla_{(\alpha} R_{\mu\nu)}=0$. The term $H_{{K}}$ represents the remaining solutions to $\nabla_{(\alpha} R_{\mu\nu)}=0$ which in general contributes both to $\nabla_\alpha R_{\mu\nu}$ and $R_{11}$. 
Our focus here in this paper will be the non-trivial CKV solutions represented by $H_{\text{Defrise}}$ and $H_{{K}}$. 
\begin{table}[ht]
    \centering
    \begin{tabular}{|l||c|c|c|c|}
    \hline 
 & $\nabla_{\mu}{\rm Riem}$ & $\nabla_{\alpha}R_{\mu\nu}$ & $\nabla_{(\alpha}R_{\mu\nu)}$ & $R_{11}$\\
 \hline\hline 
$H_{\text{Defrise}}$   &  $\neq 0$ &  $\neq 0$ & 0 & $\neq 0$\\
\hline
$H_{K}$   &  $\neq 0$ &  $\neq 0$ & 0 & $\neq 0$\\
\hline
$H_{\text{Kaigorodov}}$   &  $\neq 0$ &  $0$ & 0 & 0 \\
\hline
$H_{B}$   &  $\neq 0$ &  $0$ & 0 & 0\\
\hline
    \end{tabular}
    \caption{Generic properties of the terms included in the function $H$ considered here. }
    \label{tab:my_label}
\end{table}

Even if the CKG theory allows for a cosmological constant and thus also solutions which have accelerated expansion, it also allows for \emph{many} solutions clearly not representing our universe. Examples of these are $AdS_2\times H^2$, where $H^2$ is the hyperbolic plane; $M_2\times S^2$, where $M_2$ is Minkowski 2-space, and $S^2$ is the 2-sphere; or $AdS_3\times \mathbb{R}$. We also find generalisations of these spaces in the Kundt class of spacetimes. Once we have allowed for this 3rd order theory of gravity, the Pandora's box has been opened and oodles of solutions pour out. All of these solutions beg the question: Of all these solutions, why does the isotropic, homogeneous FLRW models with a positive cosmological constant represent our universe?

\section{Four-dimensional solutions to the CKG vacuum equations}
In the first example, we will solve for all the different terms in $H=H_{\text{Defrise}}+H_{{K}}+H_{\text{Kaigorodov}}+H_B$. However, since the first two terms are the ones that distinguishes the CKG theory most from the standard Einstein gravity, we will focus on the first two for the other solutions. We will also include the Kaigorodov term due to its geometrical significance. 

\subsection{The Siklos metrics}
The Siklos metrics were given by Siklos in \cite{Siklos}. They are a class of algebraically special solutions of the Einstein equations with a negative cosmological constant. They are conformally related to the pp-wave metrics and are often referred to as  AdS gravitational wave metrics because an interpretation of them is that they are \emph{gravitational waves propagating in an anti-de Sitter background.} These were also studied in \cite{Gurses} in the context of CKG gravity. 

We can write the class of Siklos metrics as: 
\beq
g=\frac{1}{y^2}\left[2du(dv+H(u,x,y)du)+dx^2+dy^2\right],
\eeq
which can be seen to belong to the Kundt case by the coordinate transformation $V=v/y^2$, and redefinition of the function $H$: $\bar{H}=H/y^2$. 
Choosing a null-frame
\[ \omega^1=\frac{du}{y^2},\quad \omega^2=dv+Hdu, \quad \omega^3=\frac{dx}{y}, \quad \omega^4=\frac{dy}{y},\]
and computing the Ricci tensor we get:
\[ \Ric=(\Box H)(\omega^1)^2-3[\omega^1\omega^2+\omega^2\omega^1+(\omega^3)^2+(\omega^4)^2], \quad \Box H=\frac{\partial^2H}{\partial x^2}+{y^2}\frac{\partial}{\partial y}\left(\frac{1}{y^2}\frac{\partial H}{\partial y}\right)\]
with Ricci scalar $R=-12$, i.e., constant, and hence $\nabla_\mu R=0$. These metrics are in the CSI class of Kundt metrics \cite{CSI}. 
The symmetrised Ricci tensor is therefore equal to the symmetric Harada tensor, $R_{(\mu\nu;\alpha)}=H_{\mu\nu\alpha}$. 
Assuming CKG-vacuum, $ H_{\mu\nu\alpha}=0$,  we get the equations:
\beq\label{SiklosHaradaEq}
\frac{\partial}{\partial u}(\Box H)=0, \quad \frac{\partial}{\partial x}(\Box H)=0, \quad  \frac{\partial}{\partial y}(y^4\Box H)=0
\eeq
These equations imply that $\Box H=C/y^4$. If the constant $C=0$ then we have the standard Einstein case $\Ric=-3g$. The simple non-trivial non-Einstein solution where $C\neq 0$, and thus $H_{\text{Defrise}}=c_1y^{-2}$ is the Defrise solution \cite{Defrise}. This solution was also found within the CKV theory in \cite{Gurses}. 

The general solution of eq.(\ref{SiklosHaradaEq}) can therefore be written as a sum of the Defrise term and a term satisfying a homogeneous differential equation: 
\beq
H(u,x,y)=\frac{c_1}{y^2}+F(u,x,y), 
\eeq
where $F$ is a solution of the homogeneous equation $\Box F=0$. 

Thus the equation for $F$ is the same as the Einstein requirement for $H$. This can be solved, using for example, separation of variables. Let us suppress the $u$ dependence as the equations does not depend on $u$. Then we do the separation of variables: 
\[ F(x,y)=X(x)Y(y), \] where we get the following equations: 
\beq
{X''}-\sigma^2X=0,&& \quad y^2\left(\tfrac 1{y^2}Y'\right)'+\sigma^2Y=0 \\
{X''}=0,&& \quad y^2\left(\tfrac 1{y^2}Y'\right)'=0 \\
{X''}+\sigma^2X=0,&& \quad y^2\left(\tfrac 1{y^2}Y'\right)'-\sigma^2Y=0. 
\eeq
These can be straight-forwardly solved to yield: 
\beq
F_{+,\sigma}(x,y)&=&(c_2e^{\sigma x}+c_3e^{-\sigma x})\left[c_4(\sigma y\cos\sigma y-\sin\sigma y)+c_5(\cos\sigma y+\sigma y\sin\sigma y)\right]\\
F_{0}(x,y)&=&(c_2+c_3x)(c_4y^3+1)\\
F_{-,\sigma}(x,y)&=&(c_2\cos\sigma x+c_3\sin\sigma x)\left[c_4(\sigma y-1)e^{\sigma y}+c_5(\sigma y+1)e^{-\sigma y}\right].
\eeq
All of the constants $c_2, c_3, c_4$ and $c_5$ may also depend on $u$, and the Kaigorodov term can be identified as the $c_2c_4y^3$ term. In this way, the solutions to the differential equations (\ref{SiklosHaradaEq}) are: 
\[ H(u,x,y)=\frac{c_1}{y^2}+\sum_{\sigma}F_{\bullet,\sigma}(u,x,y).\] 
The first term, $H_{\text{Defrise}}=c_1/y^2$ is the one that contains the non-trivial solution part contributing to the tensor $\nabla_\alpha R_{\mu\nu}$, the other terms, $\sum_{\sigma}F_{\bullet,\sigma}$, contribute to the $H_{\text{Kaigorodov}}+H_B$ terms.

\subsection{Solutions with AdS${}_3 \times \mathbb{R}$ backgrounds} We can also find solutions being of the following metric form: 
\beq
g=\frac{1}{y^2}\left[2du(dv+H(u,x,y)du)+dy^2\right] + dx^2
\eeq
If $H=0$ then this is the AdS${}_3 \times \mathbb{R}$ space, which is not Einstein, but is symmetric $\nabla\text{Riem}=0$. 

Also here the Defrise solution is a solution to the CKG vacuum, however, there is a bigger class of non-trivial CKG vacua: 
\[ H=\frac{1}{y^2}\left(c_1+c_2\cos 2x+c_3\sin 2x\right)+C_4(u)y^2\]
The term $H_{\text{Defrise}} =c_1/y^2$ is the Defrise term, while $H_{\text{Kaigorodov}}=C_4(u)y^2$. The term $H_K=(c_2\cos 2x+c_3\sin 2x)/y^2$ seems to be new. 

\subsection{Solutions with $M_2\times H^2$ backgrounds}
Consider the Kundt metric of the form 
\beq
g=2du(dv+H(u,x,y)du)+e^{-2qy}dx^2+dy^2,
\eeq
where $q$ is a positive constant. For $q>0$ the transverse metric is the hyperbolic plane, $H^2$. 
Here, a class of solutions can be given by
\[ H_{\sigma}(u,x,y)=(e^{\sigma x}+c_1e^{-\sigma x})e^{\frac q2y}\left[c_2J_{\alpha}\left(\tfrac{\sigma e^{qy}}q\right)+c_3Y_{\alpha}\left(\tfrac{\sigma e^{qy}}q\right)\right], \qquad \alpha=i\frac{\sqrt{7}}2,\] 
where $J_\alpha$ and $Y_\alpha$ are the Bessel functions of the 1st and 2nd kind, respectively. The constants $c_2$ and $c_3$ need to be chosen so that the $H$-function is real. Note also, that the parameter $\sigma$ can be purely imaginary in which case the $\exp(\pm\sigma x)$ should be replaced with $\sin /\cos$. Again, any sum of such solutions for different value of $\sigma$ is again a solution. 
For all of these solutions $\nabla_{\alpha}R_{\mu\nu}\neq 0$ and $R_{11}\neq 0$. In general, we get $H_K=\sum_{\sigma}H_\sigma$. 

In the special case when $\sigma=0$, then we have the simpler solutions: 
\[ H(u,x,y)=e^{\frac q2y}\left[c_1\cos\left(\frac{\sqrt{7}qy}2\right)+c_2\sin\left(\frac{\sqrt{7}qy}2\right)\right]\]
Only one of the constants $c_1$ and $c_2$ are essential. This space-time is not Einstein, nor Ricci-symmetric for $c_1,c_2\neq 0$.

\subsection{Solutions with $M_2\times S^2$ backgrounds}
Consider the Kundt metric of the form 
\beq
g=2du(dv+H(u,x,y)du)+\cos^2ydx^2+dy^2,
\eeq
where $S^2$ is the (unit) 2-sphere. Also here we can find solutions. By a separation of variables, we can find particular solutions: 
\beq
H_m(u,x,y)&=&(c_1\cos(mx)+c_2\sin(mx))\left[c_3P^m_{\ell}(\sin y)+c_4Q^m_{\ell}(\sin y)\right], \\
&&\text{where } m\in \mathbb{N}_0,\quad \ell =-\frac 12+i\frac{\sqrt{7}}2,\nonumber 
\eeq
where $P^m_\ell(x)$ and $Q^m_\ell(x)$ are the associated Legendre functions\footnote{Since the $\ell$ is complex the Legendre functions are defined using the hypergeometric functions.} of the 1st and 2nd kind, respectively. Also here we need to choose $c_3$ and $c_4$ to make sure the resulting function is real. The solutions are analogues of the $M_2\times H_2$ and possess many of the same properties. For example, $\nabla_{\alpha}R_{\mu\nu}\neq 0$, and $R_{11}\neq 0$. Clearly, these are also genuinely non-trivial CKV vacua. They all contribute to the $H_K$ term: 
\[ H_K=\sum_{m=0}^\infty a_mH_m.\]

\section{Higher-dimensional CKG vacua}
\subsection{AdS${}_m\times\mathbb{R}^n$ backgrounds}
This class of space-times can be written 
\beq\label{AdSR}
g=\frac{1}{y^2}\left[2du(dv+H(u,x^i,y,z^j)du)+(dx^1)^2+...+(dx^{m-3})^2+dy^2\right] + (dz^1)^2+...+(dz^n)^2.
\eeq
Here, the Defrise term, $H_{\text{Defrise}}=c_1/y^2$ as well as the Kaigorodov term, $H_{\text{Kaigorodov}}=C_4(u)y^{m-1}$ are again solutions. Indeed, given
\[ H=\frac{1}{y^2}\left[c_1+F(z^j)\right]+C_4(u)y^{m-1}\]
where the function $F(z_j)$ satisfies
\[ \Box_{\mathbb{R}^n} F+4F=0,\]
then the metric (\ref{AdSR}) satisfies the CKG vacuum equations. 

Straight-forwardly, a set of solutions of the above equations can be found by separation of variables:
\[ F=Z_1(z^1)Z_2(z^2)...Z_n(z^n),\]
where 
\[ Z_j''+\lambda_jZ_j=0, \qquad \sum_j\lambda_j=4.\]
\subsection{AdS$_3\times H^2$ backgrounds}
Consider the Kundt metrics of the form
\beq
g=\frac{1}{y^2}\left[2du(dv+H(u,x,y,z)du)+dy^2\right] + e^{-2qz}dx^2+dz^2.
\eeq
Again, the Defrise solutions play a role. We can find a class of solutions by defining: 
\[ H=\frac{1}{y^2}\left[c_1+F(x,z)\right]+C_4(u)y^2,\]
and let $F$ satisfy the differential equation: 
\[ \left[e^{2qz}\frac{\partial^2}{\partial x^2}+\frac{\partial^2}{\partial z^2}-q\frac{\partial}{\partial z}+(4+2q^2)\right]F=0.\]
By separation of variable, we now get the following set of solutions: 
\[ F_{\sigma}(x,z)=(e^{\sigma x}+c_2e^{-\sigma x})e^{\frac q2z}\left[c_3J_{\alpha}\left(\tfrac{\sigma e^{qz}}q\right)+c_4Y_{\alpha}\left(\tfrac{\sigma e^{qz}}q\right)\right], \qquad \alpha=i\frac{\sqrt{7q^2+16}}{2q}.\]
Again, $c_3$ and $c_4$ need to be chosen so that $F$ is real. 

Interestingly, in the case $q=\pm\sqrt{2}$, the background metric is Einstein in which case the constants $c_2$, $c_3$ and $c_4$ may also be an arbitrary function of $u$. 

In the case where $\sigma=0$, the solution simplifies to
\[ F(x,z)=e^{\frac q2z}\left[c_2\cos\left(\frac{\sqrt{7q^2+16}}2z\right)+c_3\sin\left(\frac{\sqrt{7q^2+16}}2z\right)\right]. \]

\subsection{AdS$_3\times S^2$ backgrounds}
Consider the Kundt metrics of the form
\beq
g=\frac{1}{y^2}\left[2du(dv+H(u,x,y,z)du)+dy^2\right] + a^2(\cos^2 zdx^2+dz^2).
\eeq
We can find a class of solutions by defining: 
\[ H=\frac{1}{y^2}\left[c_1+F(x,z)\right]+C_4(u)y^2.\]

By separation of variable, we now get the following set of solutions: 
\[ F_m(x,z)=(c_1\cos(mx)+c_2\sin(mx))\left[c_3P^m_{\ell}(\sin z)+c_4Q^m_{\ell}(\sin z)\right], \qquad m\in \mathbb{N}_0, \quad \ell =-\frac 12+\frac{\sqrt{16a^2-7}}2,\]
where $c_3$ and $c_4$ need to be chosen so that $F$ is real. These solutions are similar to the $M_2\times S^2$ found earlier in dimension 4. 

In the special case $a^2=1/2$, this simplifies to:
\[ F_m(x,z)=(c_1\cos(mx)+c_2\sin(mx))\left[\frac{c_3}{(\sec z+\tan z)^m}+c_4(\sec z+\tan z)^m\right].\]
If, addition, $m=0$ then: 
\[ F_0(x,z)=c_2+c_3\ln(\sec z+\tan z). \] 

\subsection{Solvegeometry gravitational wave solutions}
The Kundt metrics considered so far is a part of a larger class of Kundt gravitational wave metrics with solvmanifold backgrounds \cite{solvm2}. In particular, the Siklos metric is a special case of the solvegeometry gravitational wave class of Einstein metrics. By considering this larger class one can find more examples of non-trivial CKG vacua. 

Consider for example the 7-dimensional Einstein solvegeometry metric: 
\[ g_{7D}=2e^{-3qw}du(dv+Hdu)+e^{-4qw}(dx + ydz-zdy)^2+e^{-2qw}(dy^2+dz^2)+e^{-3qw}dX^2+dw^2, \] where $q={2\sqrt{34}}/{17}$. Here, the function $H$ can depend on all variables, except for $v$ and has to fulfill a Laplace-type equation for the space to be Einstein. 

If we assume that $H=C\exp(pw)$, then there are three choices of $p$ which makes it a CKG vacuum: 
\begin{enumerate}
    \item $p=0$: This is a Standard Einstein solvmanifold case and is a CKG vacuum simply because $R_{\mu\nu}=\lambda g_{\mu\nu}$.
    \item $p=17q/2$: This is the Kaigorodov case, which is also Einstein: $R_{\mu\nu}=\lambda g_{\mu\nu}$.
    \item $p=-3q$: This is the Defrise case which is not Einstein, has $\nabla_{\alpha}R_{\mu\nu}\neq 0$, but is a homogeneous space. This is the generalisation of the Defriese solutions which showed up earlier. 
\end{enumerate}
The authors have also checked other solvmanifolds and it seems as if \emph{for any solvegeometry gravitational waves, the non-Einstein generalised Defrise spacetime is a CKG vacuum.}

We also checked if some of the other solutions found earlier generalised to the more general class. For example, for the 8-dimensional metric: 
\[ g_{8D,\#1}=2e^{-3qw}du(dv+Hdu)+e^{-4qw}(dx + ydz-zdy)^2+e^{-2qw}(dy^2+dz^2)+e^{-3qw}dX^2+dw^2 + dY^2, \] where $q={2\sqrt{34}}/{17}$, which is simply $g_{8D}=g_{7D}+dY^2$, we found the following non-trivial CKG vacuum: 
\[ H=e^{-3qw}\left[c_1+c_2\cos(3qY)+c_3\sin(3qY)\right]+C_4(u)e^{17qw/2}.\]
Again we recognise the Defrise term, $H_{\text{Defrise}}=c_1e^{-3qw}$ as well as the Kaigorodov term, $H_{\text{Kaigorodov}}=C_4(u)e^{17qw/2}$. 
This shows that there are similar solutions as the one found in the 4-dimensional case AdS${}_3\times\mathbb{R}$. They seem to genereralise to the higher-dimensional solvegeometry gravitational waves. 

We can also check other solutions with higher-dimensional solvmanifold. Consider, for example, the 6-dimensional solvmanifold times 2-dim hyperbolic space: 
\[ g_{8D,\#2}=2e^{-3qw}du(dv+Hdu)+e^{-4qw}(dx + ydz-zdy)^2+e^{-2qw}(dy^2+dz^2)+dw^2 + e^{-2rY}dX^2+dY^2, \] where $q={2\sqrt{7}}/{7}$. Here, we again get solutions involving the Bessel functions: 
\[ H=e^{-3qw}\left[c_1+F(X,Y)\right]+C_4(u)e^{7qw},\]
where  $F$  is a sum of functions of the form 
\[ F_{\sigma}(X,Y)=(e^{\sigma X}+c_2e^{-\sigma X})e^{\frac r2Y}\left[c_3J_{\alpha}\left(\tfrac{\sigma e^{rY}}r\right)+c_4Y_{\alpha}\left(\tfrac{\sigma e^{rY}}r\right)\right], \qquad \alpha=i\frac{\sqrt{343r^2+1008}}{14r}.\]
Again, $c_3$ and $c_4$ need to be chosen so that $F$ is real. Again, $\sigma$ could be imaginary, in which case $e^{\pm\sigma X}$ should be replaced with $\cos(\sigma X)$ and $\sin(\sigma X)$. 

In the case $\sigma =0$ then 
\[  F(X,Y)=e^{\frac r2Y}\left[c_2\cos\left(\frac{\sqrt{343r^2+1008}}{14}Y\right)+c_3\sin\left(\frac{\sqrt{343r^2+1008}}{14}Y\right)\right].\]

Also solutions with a solvegeometry times the 2-sphere is possible to find. These are analogues of the one found earlier of type AdS${}_3\times H^2$. 

\section{Discussion}
In this paper, we have opened the Pandora's box of CKG theory and have shown that even in the vacuum case, the CKG theory has a myriad of different solutions. We have found 4-dimensional solutions of Kundt type and shown that in the higher-dimensional theory they are even more numerous. Many of the solutions found clearly do not represent our universe and thus a mechanism is needed to explain the current state of the universe in this theory. This can be, for example, if the current universe is an attractor solution like the de Sitter universe is in standard Einstein gravity. Or, if there is some other physical selection mechanism that prefers the more physically realistic universe models. 
More work here is clearly needed.

\end{document}